\documentclass[journal]{IEEEtran}

\usepackage{color}
\usepackage[numbers]{natbib}
\usepackage{hyperref}
\usepackage{graphicx}
\usepackage{amsmath}
\usepackage{mathtools}
\usepackage{kbordermatrix}
\usepackage[none]{hyphenat}

\ifCLASSINFOpdf
\else
\fi

\begin{document}
% paper title
% Titles are generally capitalized except for words such as a, an, and, as,
% at, but, by, for, in, nor, of, on, or, the, to and up, which are usually
% not capitalized unless they are the first or last word of the title.
% Linebreaks \\ can be used within to get better formatting as desired.
% Do not put math or special symbols in the title.
\title{Modeling the Trade-off between Throughput and Reliability in a Bluetooth Low Energy Connection}

\author{Bozheng~Pang,~\IEEEmembership{Graduate Student Member,~IEEE,}
        Tim~Claeys,~\IEEEmembership{Member,~IEEE,}
        Hans~Hallez,~\IEEEmembership{Member,~IEEE,}
        and~Jeroen~Boydens,~\IEEEmembership{Member,~IEEE}% <-this % stops a space
\thanks{Bozheng Pang, Hans Hallez, and Jeroen Boydens are with imec.DistriNet, M-Group, KU Leuven, Belgium. Tim Claeys is with ESAT-WaveCore, M-Group, KU Leuven, Belgium.}% <-this % stops a space
%\thanks{J. Doe and J. Doe are with Anonymous University.}% <-this % stops a space
\thanks{Manuscript received MM DD, 2022; revised MM DD, YYYY.}}

% The paper headers
\markboth{Journal of \LaTeX\ Class Files,~Vol.~14, No.~8, August~2015}%
{Shell \MakeLowercase{\textit{et al.}}: Bare Demo of IEEEtran.cls for IEEE Journals}

% make the title area
\maketitle

% As a general rule, do not put math, special symbols or citations
% in the abstract or keywords.
\begin{abstract}
The use of Bluetooth Low Energy in low-range Internet of Things systems is growing exponentially. Similar to other wireless communication protocols, throughput and reliability are two key performance metrics in Bluetooth Low Energy communications. However, electromagnetic interference from various sources can heavily affect the performance of wireless devices, leading to dropped throughput and unreliable communication. Therefore, there is a need for both theoretical and practical studies capable of quantifying the BLE communication performance, e.g. throughput and reliability, subject to interference. In this paper, a mathematical model to predict throughput of a BLE connection under interference is derived first, and linked to the reliability model we developed in~\cite{pang_novel_2023}. After that, extensive practical experiments are performed in various scenarios to sufficiently validate the theoretical results from both models. Finally, the trade-off between throughput and reliability is investigated through the validated models to give some inside properties of BLE communications. The similarity between the theoretical results and the experimental ones highlights the accuracy of the proposed throughput and reliability models. Hence, the two models can be used to explore the performance of various BLE designs or deployments from diverse perspectives.
\end{abstract}

% Note that keywords are not normally used for peerreview papers.

\IEEEpeerreviewmaketitle

\section{Introduction}
\label{Introduction}
\IEEEPARstart{T}{he} feasibility and excellence of Internet of Things~(IoT) have been shown in several aspects, such as industry, logistics, and smart homes~\cite{golpira_review_2021, laghari_review_2022, malik_industrial_2021}. The IoT concept arises due to the revolution of multiple technologies, and one of them is wireless communication~\cite{bhoyar_communication_2019}. For various IoT applications, different wireless communication protocols exist~\cite{singh_wireless_2021}. Bluetooth Low Energy~(BLE) is one of the most popular wireless protocols, aiming at low range and energy efficient IoT systems~\cite{furst_evaluating_2018, bluetooth_sig_bluetooth_2021}. It works in the 2.4~GHz frequency band, i.e. the industrial, scientific and medical~(ISM) band~\cite{wikipedia_ism_2022}. There are diverse wireless technologies existing in this frequency band, e.g. \mbox{Wi-Fi}, Bluetooth, and BLE itself~\cite{la_dense_2018}. As a result, BLE always faces interference challenges, for instance, due to other neighboring BLE communications~\cite{pang_comparative_2019, pang_study_2020}. The impact of interference on BLE communications has been shown in research, such as causing transmission failure and thereby degrading throughput and reliability~\cite{omar_al_kalaa_evaluating_2016, dian_formulation_2020, pang_bluetooth_2021}.

Some practical studies on BLE throughput have been conducted in~\cite{dian_practical_2018, ayoub_throughput_2020}. These two studies focused on the maximum throughput of BLE under the condition that the wireless link is error free. However, both of their results showed that the measured throughput can only achieve 94\% to 97\% of the theoretical value. After that, \citeauthor{dian_formulation_2020} introduced their study on the formulation of BLE throughput based on node and link parameters~\cite{dian_formulation_2020}. In their work, a novel scheme to formulate the throughput and the average number of successfully transmitted packets is proposed. In their proposed scheme, a prone-to-errors wireless link is modeled. This novel scheme has only been verified through simulation, while in their previous work the difference between theory, simulation, and practice has been reported.

Different from the research work mentioned above, \citeauthor{rondon_analytical_2016} published their studies on the BLE latency~\cite{rondon_analytical_2016, rondon_evaluating_2017}. In~\cite{rondon_analytical_2016}, they introduced their analytical model of the delay performance of BLE for connection-oriented applications under different bit error rate conditions. The analytical model is based on Markov chain, and the results highlighted the impact of the device's processing speed and the timing configuration of the connection on the final measured latency~\cite{agbinya_1_2019}. However, similar to~\cite{dian_formulation_2020}, the model is only validated by simulation results. Based on the analytical model published in~\cite{rondon_analytical_2016}, \citeauthor{rondon_evaluating_2017} further evaluated BLE suitability for time-critical industrial IoT applications~\cite{rondon_evaluating_2017}. Three retransmission schemes on the reliability and timeliness performance are thoroughly studied in their work, but are only evaluated by simulation results. They conclude that by optimally modifying the BLE retransmission model, a maximum delay below 46~ms and a packet loss rate in the order of $10^{-5}$ can be obtained. It is evident that their work concentrates on BLE latency instead of throughput, but the research idea of this paper is partially inspired by them, thus their work is discussed here.

Apart from throughput, reliability is also of interest in BLE communications. Hence, we conducted and proposed some research and improvements. First of all, adaptive frequency hopping~(AFH) is a scheme implemented by BLE to avoid interference~\cite{bluetooth_sig_bluetooth_2021}. Inside the AFH, two channel selection algorithms~(CSAs) are defined to help a BLE connection stay connected while hopping pseudo-randomly within the 2.4~GHz frequency band~\cite{pang_comparative_2019}. Both CSAs have been proved lack of efficiency or effectiveness under the environment with interference~\cite{pang_study_2020}. Hence, some improvements for BLE reliability were proposed~\cite{pang_bluetooth_2021, poirot_eafh_2022}. Besides, further evaluation on the proposed improvements was also conducted~\cite{pang_probability-based_2021}. Most importantly, the first BLE reliability model under the interference from other BLE connections has been developed by us~\cite{pang_novel_2023}. Our mathematical model for BLE reliability developed in~\cite{pang_novel_2023} clearly demonstrates and quantifies the impact of various BLE transmission parameters on the BLE reliability. Furthermore, the reliability model has been proved by extensive practical experiments, instead of just theory or simulation.

According to all the literature discussed above, the throughput and reliability of BLE have been investigated from different perspectives. However, there is no deeper research showing the relation or the trade-off between these two communication performance metrics. Hence, in this paper, the trade-off between throughput and reliability is thoroughly investigated, i.e. modeled and experimentally validated. In order to provide a thorough understanding of the relationship between the communication performance and its various connection parameters, multiple BLE communication parameters are involved, such as packet length and number of packets. In other words, this paper can serve as a design-level guideline for BLE usage or deployment.

It is worth mentioning that, to the best of our knowledge, this paper is the first one that describes a thorough model of the trade-off between BLE throughput and reliability, and validatesall the results by practical experiments. There are no existing approaches that can accurately quantify the trade-off or relationship yet. The impact of each connection parameter on the throughput and reliability can be clearly illustrated by the model introduced in this paper. The significance of the suggested model is to explain the trade-off within BLE communications simply through numbers and formulas. Rather than just providing a rough or general trend, the trade-off and relationships between reliability and throughput are accurately calculated under various scenarios. The model is highly recommended for BLE users/developers to better deploy their BLE devices, e.g., a beforehand design, a straightforward control, and convenient management on the BLE communication and network. Note that although it is unrealistic to resolve all the throughput and reliability challenges just by \mbox{fine-tuning} BLE connection parameters, this research can be considered as a stepping stone for further steps. The contributions of this paper are summarized as follows:

\begin{enumerate}
	\item A mathematical model to quantify the throughput of a BLE connection under interference is derived and optimized. It is considered a useful tool to estimate the BLE throughput with different combinations of BLE parameters under different interference environments. The derived throughput model is linked to the reliability model we developed in~\cite{pang_novel_2023}. Using these two models, we will discuss the trade-off between throughput and reliability in a quantitative way.
	\item All the theoretical results from both the throughput model and the reliability model are validated by extensive practical experiments. Various sets of BLE connection parameters and interference environments or scenarios are applied to verify different aspects of the proposed models. Note that the details of the reliability model and related validation experiments can be found in~\cite{pang_novel_2023}, hence, this paper focuses more on the throughput model.
	\item The trade-off between throughput and reliability within BLE communications is investigated through the two validated models. Several Pareto curves between throughput and reliability under different scenarios are drawn to visualize the compromise within BLE communications under interference~\cite{blanchet_applying_2018,blanchet_generalized_2022}.
\end{enumerate}

The remainder of the paper is organized as follows. In Section~\ref{Mathematical Models}, the background of BLE communications is first introduced briefly, to help the readers get a better understanding of the rest of the paper. After that, two mathematical models and the link between them are introduced. In Section~\ref{Experimental Setup}, a description of the experimental setup used to prove the theoretical results is shown. The comparison between the models and the experiments, and a discussion on the trade-off between throughput and reliability are provided in Section~\ref{Results and Discussion}. In Section~\ref{Conclusion}, this paper is concluded with some final remarks and possible future work.

\section{Mathematical Models}
\label{Mathematical Models}
This section first presents some necessary background knowledge of BLE communications. Then the two mathematical models, namely the throughput model and the reliability model, are derived. The throughput model is derived in detail due to its novelty, while the reliability model is only introduced briefly since it has been derived and validated in detail in our previous work~\cite{pang_novel_2023}. Finally, the link of the two models is described briefly.

\subsection{Background}
BLE is a wireless personal area network technology designed for novel applications, such as healthcare, fitness, smart home, and industries~\cite{yuan_embracing_2021, porjazoski_bluetooth_2019, magadan_low-cost_2020}. It supports different communication modes, e.g. connectionless and connection-oriented. The connection-oriented mode is the focus of this paper since it is designed more for data exchange comparing with the connectionless mode~\cite{garcia-ortiz_experimental_2021, siva_connection-oriented_2020}. In the connection-oriented mode, at least two BLE devices are used to create a BLE connection. One of the devices is defined as a central, and the other one as a peripheral~\cite{bluetooth_sig_bluetooth_2021}.

As mentioned before, BLE is widely used in the 2.4~GHz frequency band. The spectrum usage of BLE is managed by the two CSAs defined in the BLE specification~\cite{bluetooth_sig_bluetooth_2021}. The 2.4~GHz frequency band is divided into 37 data channels and the CSAs are used to calculate a pseudo-random channel for the BLE connection~\cite{pang_comparative_2019}.

Each time a BLE connection hops to a channel, it stays there for a certain amount of time. This is called a connection interval~\cite{bluetooth_sig_bluetooth_2021}. The value of the connection interval is negotiated between the central and the peripheral at the beginning of the connection. The data exchange of the central and peripheral occurs at the start of each connection interval. It is always initialized by a packet sent by the central, and followed by a packet from the peripheral. This data transaction can be repeated numerous times during a single connection interval, depending on the amount of data to be sent. A connection event is made up of all the transactions that occur inside the same connection interval.

\subsection{Throughput Model}
Besides basic knowledge about BLE communications, the retransmission mechanism of BLE needs to be mentioned. Without interference, a BLE connection is able to follow the communication process described above. However, the connection is different under interference, since the BLE retransmission mechanism is activated when faced with interference~\cite{bluetooth_sig_bluetooth_2021, ortiz_evaluation_2021}.

Under interference, the packets exchanged between the central and the peripheral might experience situations like packet corruption and packet loss~\cite{tipparaju_mitigation_2021, kalinin_iot_2021}. When invalid packets occur, a retransmission is necessary. According to the BLE specification, the number of retransmissions for a packet is unlimited~\cite{bluetooth_sig_bluetooth_2021}. It suggests that an invalid packet is retransmitted until it is correctly received and acknowledged. However, that may lead to an infinite number of attempts, thus there are some basic rules defined in the BLE specification to limit some aspects of the retransmission:
\begin{enumerate}
\item A successful transaction counts when both packets from the central and the peripheral are valid, suggesting no bit errors. In this case, the transaction state is \textit{success}.

\item If bit errors appear in a packet but not in the access address, i.e. invalid cyclic redundancy check~(CRC), a transmission failure is counted, and a retransmission is required. The connection event remains open, and the retransmission is immediately performed at the next transaction. In this case, the transaction state is defined as \textit{fail~(open)}.

\item If the bit errors appear in the access address of a packet, a transmission failure is counted, and a retransmission is required. However, the connection event is immediately terminated, thus the required retransmission could only occur in the next connection interval. In such a case, the transaction state is called \textit{fail~(close)}.

\item Two consecutive packets received with an invalid CRC match within a single connection event shall close the event. And they will be retransmitted in the next connection interval. In this situation, the transaction state is also \textit{fail~(close)}.
\end{enumerate}

With all the retransmission rules introduced, the throughput model is derived. It is based on the Markov chain~\cite{agbinya_1_2019}. The goal of this mathematical model is to predict the BLE throughput under various conditions. For instance, different interference strengths and diverse combinations of BLE connection parameters.

% TODO: \usepackage{graphicx} required
\begin{figure}
	\centering
	\includegraphics[width=1\linewidth]{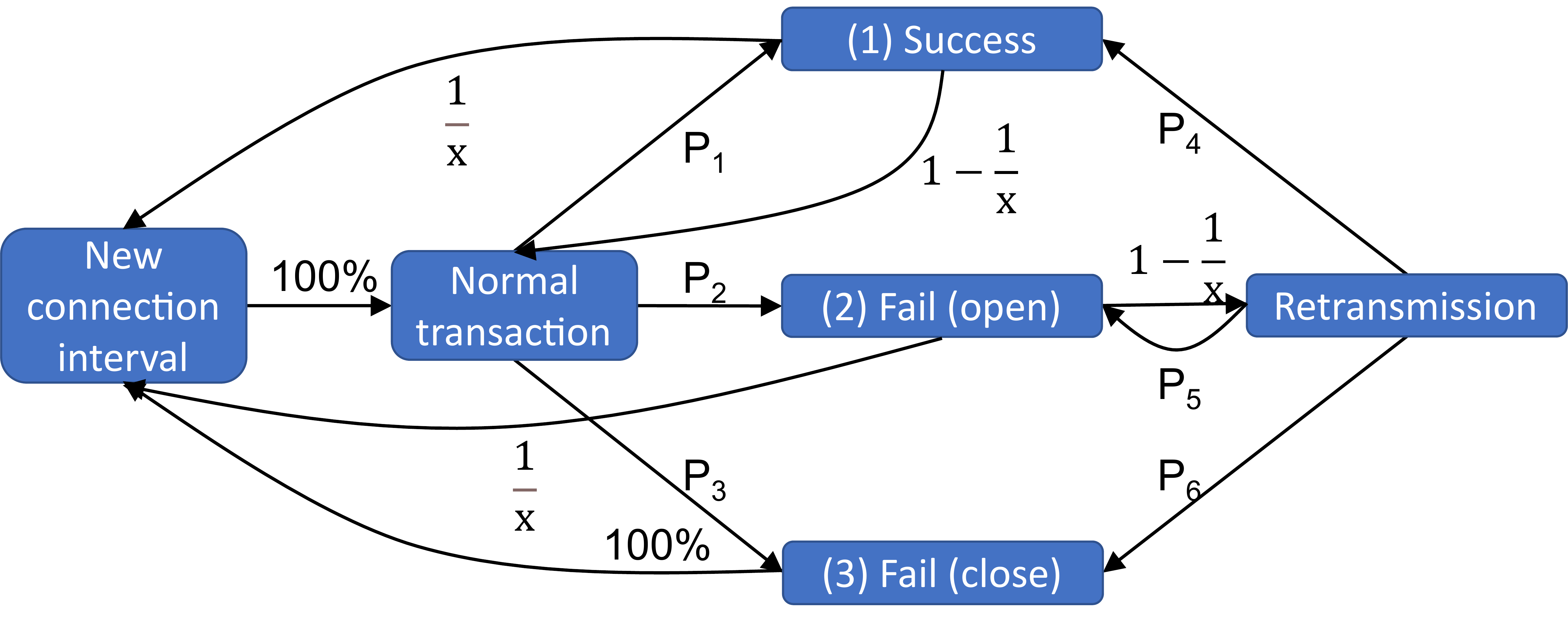}
	\caption{Graphical representation of possible status of transactions in BLE communications, in the form of a Markov chain.}
	\label{fig:fig01}
\end{figure}
First of all, a graphical representation of the BLE communication details is shown in Fig.~\ref{fig:fig01} through the form of a Markov chain. It summarizes all the possible events for a BLE transaction and represents the BLE communication in reality. Roughly speaking, a BLE transaction is sent at the beginning of a connection interval as a \textit{normal transaction}. So the probability from the event \textit{new connection interval} to \textit{normal transaction} is always 100\%. Then there are three possible states for the transaction, which are \textit{success}, \textit{fail~(open)}, and \textit{fail~(close)}.

The \textit{success} state represents that a transaction is without any bit errors, corresponding to the basic rule 1 of the retransmission scheme. The probability of a successful transmission is indicated as $P_1$. After a \textit{success}, there are two options for the BLE connection: (1) starting another \textit{new connection interval}, and (2) continuing with another \textit{normal transaction} in the same connection interval. It depends on the number of transactions~(x) arranged/allowed in the connection interval. For instance, when only one transaction is allowed in the connection interval, the BLE connection must start a \textit{new connection interval} to send the next transaction. The $\frac{1}{x}$ on the left top of Fig.~\ref{fig:fig01} represents the probability from the \textit{success} status to the \textit{new connection interval}. Consequently, another \textit{normal transaction} in the same connection interval is not continued since the probability of that process is 0\%~($x=1\rightarrow1-\frac{1}{x}=0\%$).

The \textit{fail~(open)} status suggests that there are bit errors in the transaction but not in the access address, corresponding to the basic rule 2 of the retransmission scheme. The BLE transaction goes into \textit{fail~(open)} with a probability defined as $P_2$. In this case, the \textit{retransmission} has a probability of $1-\frac{1}{x}$. Similar to the situation discussed above, it depends on the number of transactions arranged in the same connection interval. The \textit{fail~(open)} status is the only chance for the BLE transaction to go into the \textit{retransmission} status. Similar to the \textit{normal transaction}, there are three possible status for a \textit{retransmission} transaction, i.e. \textit{success}, \textit{fail~(open)}, and \textit{fail~(close)}, with $P_4$, $P_5$, and $P_6$ as their probabilities respectively.

The \textit{fail~(close)} means that bit errors occur to the transaction and lead to the connection event to be closed, corresponding to the basic rules 3 and 4 of the retransmission scheme. A \textit{normal transaction} ends up with this status with a probability of $P_3$. While a \textit{retransmission} transaction has a probability of $P_6$ to close the current connection event. When the transaction is in the \textit{fail~(close)} stage, it results in a \textit{new connection interval} with a 100\% probability.

\begin{figure}
	\centering
	\includegraphics[width=1\linewidth]{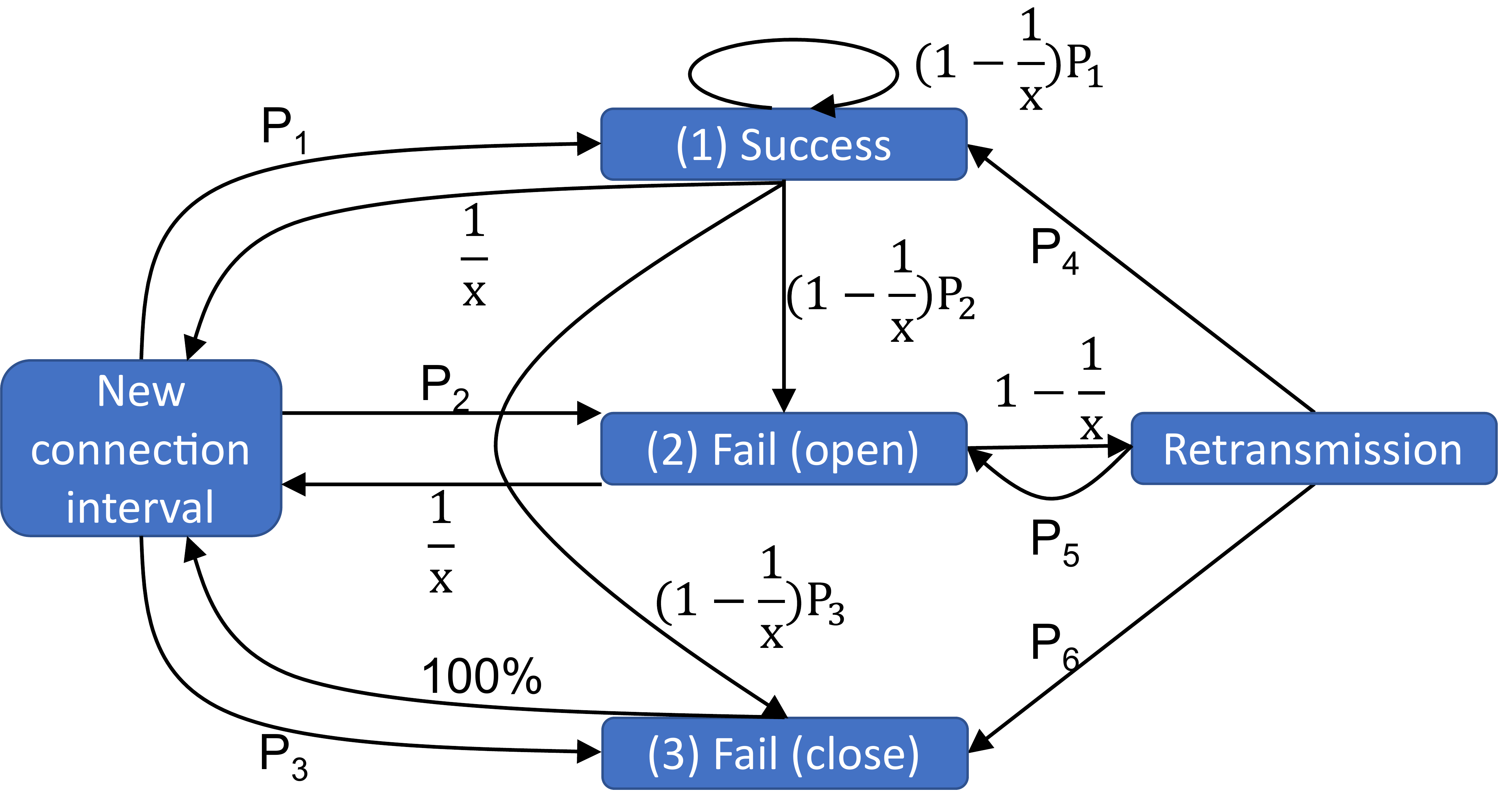}
	\caption{One step of Markov chain simplification by eliminating the status normal transaction in Fig.~\ref{fig:fig01}.}
	\label{fig:fig02}
\end{figure}
To model the throughput under interference, we are mostly interested in the three transaction states, namely \textit{success}, \textit{fail~(open)}, and \textit{fail~(close)}. Hence, the Markov chain needs to be optimized or simplified. The purpose of the optimization and simplification is to reduce the amount of information in the Markov chain, leaving only the necessary information. In general, three states can be simplified from the Markov chain, i.e. \textit{new connection interval}, \textit{normal transaction}, and \textit{retransmission}. As an example, the principle and steps to simplify the \textit{normal transaction} status are shown in Fig.~\ref{fig:fig02} and explained in detail below.

As mentioned in Fig.~\ref{fig:fig01}, the \textit{new connection interval} leads to a \textit{normal transaction} with a probability of 100\%. A \textit{normal transaction} has a probability of $P_1$ to be successfully transmitted. Hence, the probability between the \textit{new connection interval} and the \textit{success} stage is $100\% \times P_1$~(see Fig.~\ref{fig:fig02}). Similar results of the probabilities between the \textit{new connection interval} and the \textit{fail~(open)} and the \textit{fail~(close)} can be calculated as $P_2$ and $P_3$ respectively~(see Fig.~\ref{fig:fig02}). By checking Fig.~\ref{fig:fig01}, from the \textit{success} status, a chance exists to go back to the \textit{normal transaction}. Therefore, another three probabilities need to be calculated. These are the probabilities between \textit{success} and \textit{success}, \textit{success} and \textit{fail~(open)}, and \textit{success} and \textit{fail~(close)}. Similarly, according to Fig.~\ref{fig:fig01}, the \textit{success} goes to the \textit{normal transaction} with a probability of $1-\frac{1}{x}$, while the \textit{normal transaction} goes back to the \textit{success} stage with a probability of $P_1$. As a result, the probability of \textit{success} turning into itself is $(1-\frac{1}{x}) \times P_1$. Under the same logic, the probability results between \textit{success} and \textit{fail~(open)}, and \textit{success} and \textit{fail~(close)}, are calculated as $(1-\frac{1}{x}) \times P_2$, and $(1-\frac{1}{x}) \times P_3$ respectively.

\begin{figure}
	\centering
	\includegraphics[width=1\linewidth]{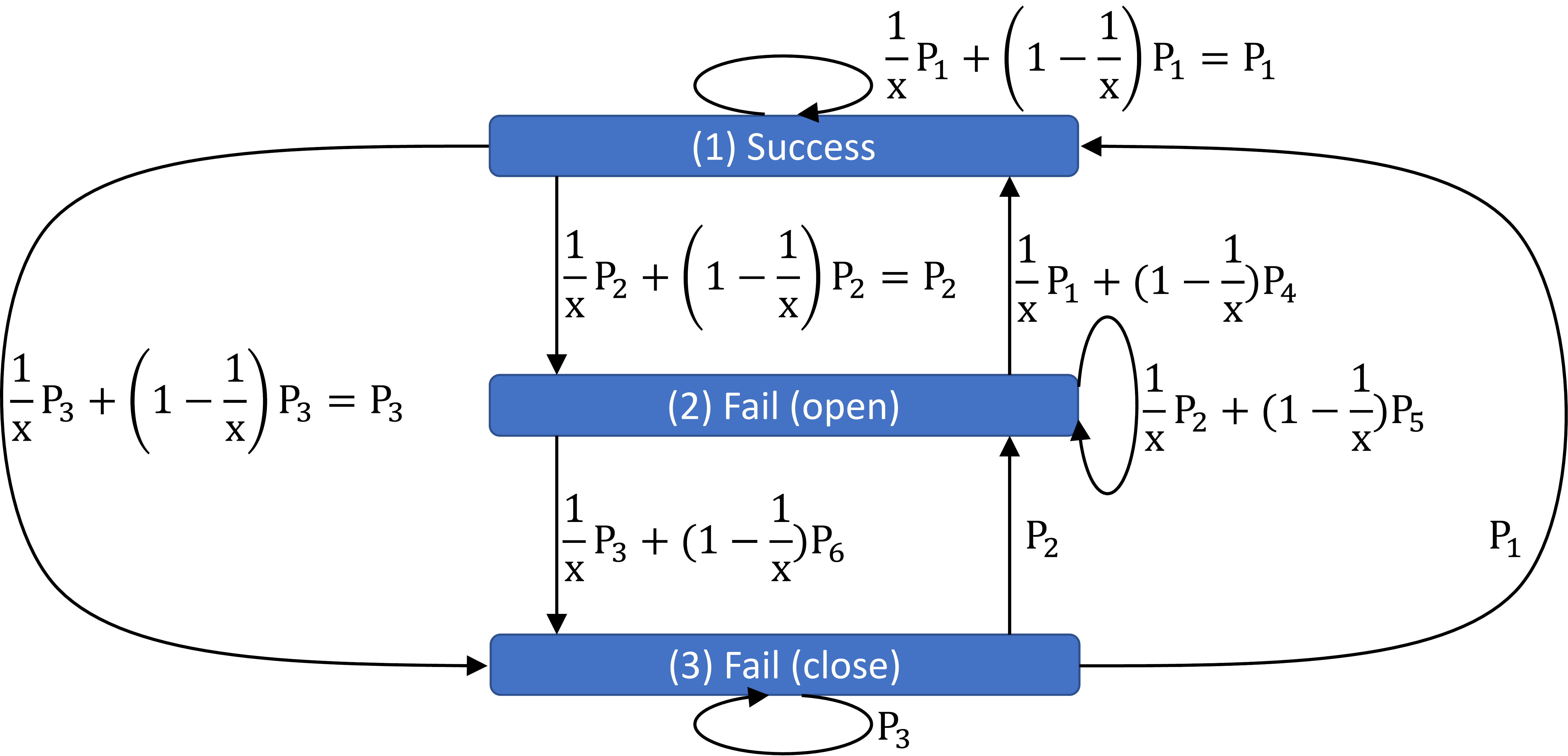}
	\caption{Final simplified version of the Markov chain to represent the three possible outcomes of a single transaction. The three possible outcomes are (1)~Success, (2)~Fail~(open), and (3)~Fail~(close).}
	\label{fig:fig03}
\end{figure}
Following the same principle and steps explained above, the Markov chain is finally simplified into Fig.~\ref{fig:fig03}. Where only three stages are left, i.e. \textit{success}, \textit{fail~(open)}, and \textit{fail~(close)}, and the transition probabilities among them are listed in the graph. With this simplified Markov chain of BLE communication, the transition matrix~($A$) can be written as follows:
\begin{equation*}
	\scriptsize
	A = 
	\kbordermatrix{
		& (1) & (2) & (3) \\
		(1) &
			P_1 & P_2 & P_3 \\
		(2) &
			\frac{1}{x}P_1+(1-\frac{1}{x})P_4 & 
			\frac{1}{x}P_2+(1-\frac{1}{x})P_5 & 
			\frac{1}{x}P_3+(1-\frac{1}{x})P_6 \\
		(3) &
			P_1 & P_2 & P_3
	}
\end{equation*}

The transition matrix $A$ is a $3\times3$ matrix that illustrates the transition probabilities between either two status in the Markov chain of Fig.~\ref{fig:fig03}~\cite{wikipedia_markov_2022}. For example, $A_{(2)(3)}$ represents the transition probability from status (2) \textit{fail~(open)} to status (3) \textit{fail~(close)}, which equals $\frac{1}{x}P_3+(1-\frac{1}{x})P_6$.

According to the property of a Markov chain, given an initial distribution~($\pi_0$), after a sufficiently long time, e.g. $n$ generations/iterations, the generation $n$~($\pi_n$) does not change any longer, which is called a stationary distribution~\cite{roy_convergence_2019}. According to the accuracy requirement of the applications, it can be decided when the iterations should stop, e.g., when the third decimal place stays stationary. The mathematical representation of this property is:
\begin{equation}
	\small
	\label{equation01}
	\begin{aligned}
		\pi_n	& = \pi_{n-1} \times A=\pi_0 \times A^n \\
				& = \begin{bmatrix}
					Num_{Success} & Num_{Fail~(open)} & Num_{Fail~(close)}
					\end{bmatrix}
	\end{aligned}
\end{equation}

$\pi_n$ is comprised of three terms, which are $Num_{Success}$, $Num_{Fail~(open)}$, and $Num_{Fail~(close)}$. As the stationary distribution vector, the first element of $\pi_n$, which is $\pi_n[0]$ and equals to $Num_{Success}$, represents how many transactions or packets are finally successfully transmitted. The sum of all the elements in the $\pi_n$ is the total number of transmitted transactions or packets. With these two numbers, a ratio called transmission success ratio~(TSR) can be defined as:
\begin{equation}
	\small
	\label{equation02}
	\text{TSR}=\frac{\pi_n[0]}{sum(\pi_n)}
\end{equation}

The TSR defined above describes the ratio between the successful transactions/packets and the total transactions/packets. The ideal throughput is defined in Equation~\eqref{equation03}, where the $PL$ represents the BLE packet length in the unit of bytes, the $x$ is the number of transactions/packets in each connection interval, and the $CI$ represents the connection interval length in the unit of seconds. As we want to express the throughput in bits instead of bytes, a multiplication factor of 8 is added to the numerator. This idea throughput represents the throughput when all the transactions/packets are successfully transmitted, such as under an environment with no noise.
\begin{equation}
	\small
	\label{equation03}
	\text{Throughput}_i=\frac{PL \times 8 \times x}{CI}
\end{equation}

The real throughput under interference can be calculated as the product of the TSR and ideal throughput, which is shown in Equation~\eqref{equation04}. Both throughputs are calculated into the unit of bits per second~(bps).
\begin{equation}
	\small
	\label{equation04}
	\text{Throughput}_r=\text{TSR} \times \text{Throughput}_i
\end{equation}

Till now, the throughput calculation framework has been set up. With the transition matrix $A$, the stationary distribution $\pi_n$ can be found easily, thus the TSR is solved. However, currently the distribution matrix $A$ is composed of six unknown probabilities, namely $P_1$ to $P_6$. They are discussed in the following paragraphs and equations.

According to Fig.~\ref{fig:fig01}, $P_1$ to $P_6$ describe the probabilities of the \textit{normal transaction} or the \textit{retransmission} going into the three possible states, i.e. \textit{success}, \textit{fail~(open)}, and \textit{fail~(close)}. Based on literature~\cite{freschi_study_2019, pang_novel_2023}, the successful probability of transferring a series of data bits depends on the bit error rate~(BER) and bit length of the data. To describe the dependency, Equation~\eqref{equation05} is introduced.
\begin{equation}
	\small
	\label{equation05}
	\rho=(1-BER)^l
\end{equation}

Equation~\eqref{equation05} defines the successful probability $\rho$ of transferring $l$ data bits under a $BER$ defined by a specific interference condition. With this initial equation, six basic equations are defined as follows.
\begin{equation}
	\small
	\label{equation06}
	\begin{aligned}
		\rho_{AA}=(1-BER)^{l_{AA}} \\
		\rho_{CP}=(1-BER)^{l_{CP}-l_{AA}} \\
		\rho_{PC}=(1-BER)^{l_{PC}-l_{AA}} \\		
		q_{AA}=1-\rho_{AA} \\
		q_{CP}=1-\rho_{CP} \\
		q_{PC}=1-\rho_{PC}
	\end{aligned}
\end{equation}

Based on the four basic rules of BLE communication, six basic equations are defined in~\eqref{equation06}. $AA$ is the abbreviation of access address. $CP$ and $PC$ represent the directions of the packets: from central to peripheral and peripheral to central respectively. $l_{AA}$ is the bit length of the access address, which is 32~bits in BLE. $l_{CP}$ and $l_{PC}$ are the number of bits of the packets from different directions. $l_{CP}-l_{AA}$ and $l_{PC}-l_{AA}$ are the numbers of bits in the packets except the access address. As mentioned before, $\rho$ represents the probability of a successful transmission of a certain amount of data bits, while $q$ represents the probability if the transmission of at least one bit in the packet is unsuccessful. Thus $\rho_{AA}$ refers to the success probability of the access address within a BLE packet. $\rho_{CP}$ and $\rho_{PC}$ denote the success probability of the other data bits within the BLE packet except the access address. On the contrary, $q_{AA}$, $q_{CP}$, and $q_{PC}$ are the failure probabilities respectively.

With all the probabilities defined for different parts of the BLE packet, the probabilities in the Markov chain of Fig.~\ref{fig:fig01}, i.e. $P_1$ to $P_6$, can be derived. First of all, $P_1$ is the success probability of a \textit{normal transaction}. To successfully transfer the \textit{normal transaction}, both packets from the central and the peripheral should be without bit errors to both the access address in the packets and all the other data bits. Note that $\rho_{AA}\rho_{CP}$ is the multiplication of $\rho_{AA}$ and $\rho_{CP}$, and represents the successful transmission probability of the central packet. It is written without $\times$ between $\rho_{AA}$ and $\rho_{CP}$ to illustrate that they both belong to a same packet. The $\times$ between $\rho_{AA}\rho_{CP}$ and $\rho_{AA}\rho_{PC}$ means that the combination of both of them results in a transaction. Hence, $P_1$ is calculated as the product of:
\begin{equation}
	\small
	\label{equation07}
	P_1=\rho_{AA}\rho_{CP} \times \rho_{AA}\rho_{PC}
\end{equation}

$P_2$ is the probability of the \textit{normal transaction} going into the \textit{fail~(open)} status. The calculation idea is similar to $P_1$. Considering the four communication rules defined in BLE specification, $P_2$ can be calculated by:
\begin{equation}
	\small
	\label{equation08}
	\begin{split}
		P_2=
		\rho_{AA}q_{CP} \times \rho_{AA}\rho_{PC} \\
		+ \rho_{AA}\rho_{CP} \times \rho_{AA}q_{PC} \\
		+ \rho_{AA}q_{CP} \times \rho_{AA}q_{PC}
	\end{split}
\end{equation}

There are three terms in Equation~\eqref{equation08}, which represent three possible cases between the \textit{normal transaction} and the \textit{fail~(open)} status. The first term suggests the case that at least one bit error occurs only in the packet sent from the central, and not in the access address domain. The second term gives the probability that at least one bit error occurs only in the packet sent from peripheral, and not in the access address domain. While the last term means that some bit errors are in the packets from both the central and the peripheral, and not in the access address domains. Following the same principle and based on the BLE communication rules, $P_3$ is the probability of bit errors occurring in the access address domains of either packet or both. The mathematical representation is:
\begin{equation}
	\small
	\label{equation09}
	P_3=q_{AA}+\rho_{AA}q_{AA}
\end{equation}

Apart from $P_{1-3}$, $P_{4-6}$ are originated from the \textit{retransmission} state. The \textit{retransmission} status can only be from the \textit{fail~(open)} status. Therefore, the calculation for $P_4$ to $P_6$ is based on $P_2$. However, to calculate the marginal probability from the \textit{retransmission} to the \textit{success}, the probability between the \textit{normal transaction} and the \textit{fail~(open)}, i.e. $P_2$, must be eliminated from $P_4$. As a result, $P_2$ is shown as the denominator of $P_4$, $P_4$ is calculated as below.
\begin{equation}
	\small
	\label{equation10}
	P_4=
	\frac
	{(\rho_{AA}q_{CP} \times \rho_{AA}q_{PC}) \times 
	(\rho_{AA}\rho_{CP} \times \rho_{AA}\rho_{PC})}
	{P_2}
\end{equation}

The calculation of Equation~\eqref{equation10} is similar to the successful \textit{normal transaction}. A successful \textit{retransmission} asks for both packets to be without bit errors. Besides, only the last term of Equation~\eqref{equation08} is shown in $P_4$. This is to avoid the packets from the next transaction to be involved into the current one. If they are involved, more packets are counted as successfully transmitted. This will lead to the increment of TSR, which further results in the inaccuracy of the throughput model.

$P_5$ is the probability of the \textit{retransmission} turning back to the \textit{fail~(open)} stage. To achieve this, there must be some bit errors existing in the retransmission, meanwhile those bit errors do not lead to the connection interval to be terminated. It suggests that the bit errors do not exist in the access address of the retransmitted packets, or the same packet which leads the \textit{normal transaction} into \textit{fail~(open)}, as referred to the rule 4 of BLE communication. With these limitations in mind, $P_5$ is defined in Equation~\eqref{equation11}. It is worth mentioning that, the packets from the next transaction mentioned in the last paragraph are classified as a part of $P_5$, since they can be considered neither a part of \textit{success} nor \textit{fail~(close)}. As a result, $P_5$ is written as:
\begin{equation}
	\small
	\label{equation11}
	\begin{split}
		P_5=
		\frac
		{\splitfrac{(\rho_{AA}q_{CP} \times \rho_{AA}\rho_{PC}) \times
		(\rho_{AA}\rho_{CP} \times \rho_{AA})}
		{+ (\rho_{AA}\rho_{CP} \times \rho_{AA}q_{PC}) \times
		(\rho_{AA} \times \rho_{AA}\rho_{PC})}}
		{P_2}
	\end{split}
\end{equation}

Following the same logic, $P_6$ aims to fail the \textit{retransmission} and close the current connection event with $P_2$ as the prerequisite. To not repeat the similar derivation process too much, $P_6$ is given in Equation~\eqref{equation12} without further description. Note that $P_6$ can also be calculated by $1-P_4-P_5$, which is an easier way. But here it is derived from scratch to ensure the rigor of this paper. The easier way can be considered a check for the derivation, which has been validated during our experiments.
\begin{equation}
	\scriptsize
	\label{equation12}
	\begin{split}
		P_6=
		\frac
		{
			\splitfrac{\splitfrac
			{(\rho_{AA}q_{CP} \times \rho_{AA}\rho_{PC}) \times
				((1-\rho_{AA}\rho_{CP})+\rho_{AA}\rho_{CP} \times q_{AA})}
			{+ (\rho_{AA}\rho_{CP} \times \rho_{AA}q_{PC}) \times
				(q_{AA} + \rho_{AA} \times (1 - \rho_{AA}\rho_{PC}))}}
			{+ (\rho_{AA}q_{CP} \times \rho_{AA}q_{PC}) \times
				(1-\rho_{AA}\rho_{CP} \times \rho_{AA}\rho_{PC})}
		}
		{P_2}
	\end{split}
\end{equation}

Finally, with the probabilities $P_1$ to $P_6$~(Equations~\eqref{equation07} to~\eqref{equation12}), the transition matrix $A$ can be calculated. With the matrix $A$ and a random initial distribution $\pi_0$, the stationary distribution vector $\pi_n$ can be obtained. As a result, TSR and $\text{Throughput}_r$ in Equations~\eqref{equation02} and~\ref{equation04} are achieved. Till now, the BLE throughput model is fully developed, and the $\text{Throughput}_r$ is the predicted value of the BLE throughput for a given BER. It is expected to be close to the measured throughput under the same condition and parameter settings, such as packet length and connection interval.

\subsection{Reliability Model}
After the throughput model, a BLE reliability model is introduced. The reliability model is derived and validated in detail in~\cite{pang_novel_2023}. It has been validated by extensive practical experiments in~\cite{pang_novel_2023}, and is further confirmed by some extra experiments in this paper while being combined with the proposed throughput model. The reliability model is defined with the equations below.
\begin{equation}
	\small
	\label{equation13}
	\begin{split}
		&P_{TF}
		=(1-(1-BER_V)^{2L_V}) \\
		&\times min(1,\frac{m(PT_V+IFS)+n(PT_D+IFS)}{CI_D}) \\
		&\times (1-max(0,\frac{IFS-PT_V}{PT_D+IFS})^m)
	\end{split}
\end{equation}

\begin{table}[]
	\centering
	\caption{Symbols in Equation~\eqref{equation13}}
	\label{tab:table1}
	\begin{tabular}{|l|p{7cm}|}
		\hline
		\multicolumn{1}{|c|}{\textbf{Symbol}} & \multicolumn{1}{c|}{\textbf{Definition}}
		\\ \hline \hline
		$V$ & BLE connection under interference~(victim)
		\\ \hline
		$D$ & BLE connection creating interference~(disturber)
		\\ \hline
		$BER_V$ & Bit error rate measured on the victim connection, corresponding to the $BER$ in the developed throughput model
		\\ \hline
		$L_V$ & (Average) Bit length sent in the victim connection, equivalent to $l_{CP}$ or $l_{PC}$
		\\ \hline
		$m$ & Number of packets from both central and peripheral in the victim connection, equivalent to two times the number of transactions~($2x$)
		\\ \hline
		$n$ & Number of packets from both central and peripheral in the disturber connection
		\\ \hline
		$PT_V$ & (Average) Packet transmission time of a victim packet
		\\ \hline
		$PT_D$ & (Average) Packet transmission time of a disturber packet
		\\ \hline
		$CI_D$ & Connection interval of the disturber connection
		\\ \hline
		$IFS$ & Inter frame space, equivalent to 150~$\mu$s
		\\ \hline
	\end{tabular}
\end{table}

Equation~\eqref{equation13} calculates the probability of a transmission failure for a BLE connection~($P_{TF}$). All the necessary parameters to calculate $P_{TF}$ are listed in Table~\ref{tab:table1}. As a result, the reliability of a BLE connection~(victim) under interference is written as:
\begin{equation}
	\small
	\label{equation14}
	\begin{aligned}
		Reliability=1-P_{TF} \\
	\end{aligned}
\end{equation}

In Equation~\eqref{equation14}, $Reliability$ calculates the reliability of the victim connection under the interference of the disturber connection. It is directly related to the transmission failure, which can be estimated by the packet loss rate measured in a BLE connection.

Till now, both the throughput model and the reliability model are shown. With both models in hand, the link between them can be introduced and analyzed.

\subsection{Combination and Analysis}
As can be seen from both developed models, there are several common/related parameters inside, such as BER, packet/bit length, number of transactions/packets. Hence, it is achievable to link the throughput model to the reliability model mathematically through those common parameters. For instance, by eliminating the BER in both models, we obtain a model where both throughput and reliability are present. Next to the combination, some analyses and further illustrations of the two models are mentioned below.

The correctness of the proposed throughput model can be validated by some properties of Markov chain~\cite{wikipedia_markov_2022}. In Fig.~\ref{fig:fig01}, the sum of $P_1$, $P_2$, and $P_3$ should be 100\%, same for the probabilities, $P_4$, $P_5$, and $P_6$. Similarly, in the transition matrix $A$, the sum of each line should be 100\% as well. For example, the sum of the second line of $A$ is $\frac{1}{x}(P_1+P_2+P_3)+(1-\frac{1}{x})(P_4+P_5+P_6)$. It can be further simplified as $\frac{1}{x}+(1-\frac{1}{x})=100\%$, as long as the six probabilities follow the Markov chain property mentioned before. Another instance is the stationary distribution vector $\pi_n$. The sum of all the three elements in $\pi_n$ should equal to the sum of the initial distribution $\pi_0$. With these mentioned properties, the throughput model can be easily validated during any experiments.

The reliability model has been developed and discussed in detail in~\cite{pang_novel_2023}. However, for the ease of understanding in this paper, some details are mentioned here. First, the reliability model is more applicable when the interference source is another BLE connection, since it aims to study the coexistence between multiple BLE pairs. As a result, the experiments in this paper also use BLE as the interference source. Second, although there are several common parameters/symbols in the two models, they should be taken care of when employing them. For instance, the $PT_V$ in Equation~\eqref{equation13} is the packet transmission time of a single packet from either the central or the peripheral in the victim connection. Hence, given the corresponding $PL$ to Equation~\eqref{equation03}, the throughput is calculated as a single-direction throughput. For a bidirectional throughput, the packet transmission time from the other device also should be considered.

\section{Experimental Setup}
\label{Experimental Setup}
The experimental setup introduced in this section aims to validate and illustrate the accuracy of the newly introduced throughput model and the earlier developed reliability model. After the experiments, the two models are considered trustworthy with tolerable errors, and are further utilized to discuss the trade-off between throughput and reliability in a BLE connection.

\begin{figure}
	\centering
	\includegraphics[width=1\linewidth]{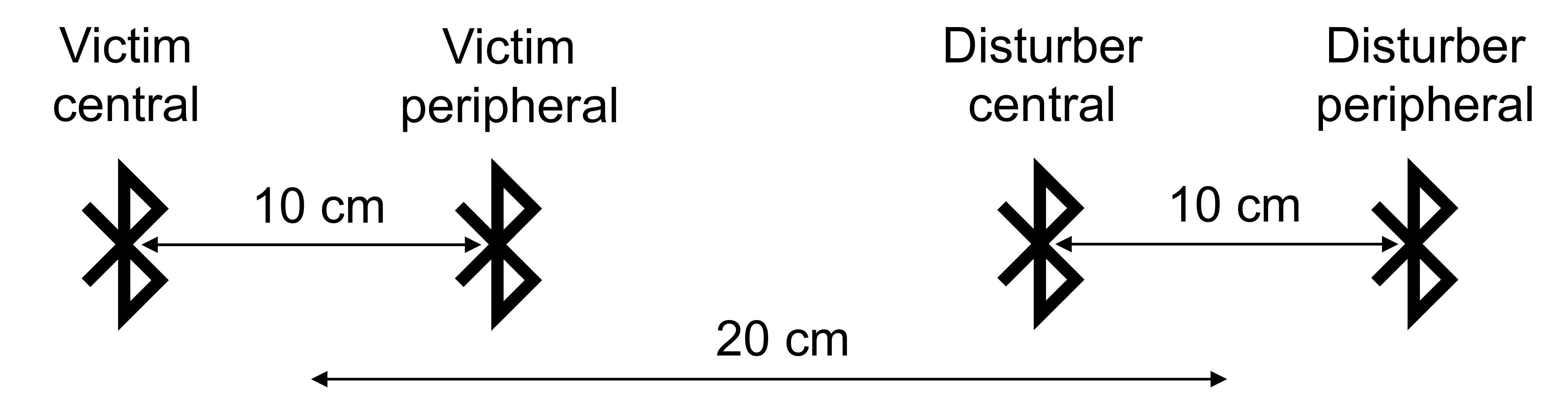}
	\caption{Schematic of the deployment of two BLE connections.}
	\label{fig:fig04}
\end{figure}
The experimental setup is similar to the one designed in~\cite{pang_novel_2023}. The object under interference is a BLE connection, and the interference source is another BLE pair. Each BLE connection is built up using two nRF52840 DK development boards, placed in a noiseless office environment~\cite{nordic_semiconductor_nrf52840_2022}. Various scenarios can be achieved by differing parameters, such as transmission power and connection event structure. In this paper, three specific scenarios are tested and discussed. Since each scenario represents a different electromagnetic environment, the developed models can be considered widely examined and validated. The deployment of the two BLE connections are graphically shown in Fig.~\ref{fig:fig04}. It is worth noting that, although the location of the BLE devices is constant, the variation of other parameters is sufficient to simulate diverse interference environments. For example, the amount of data exchanged in the disturber connection.

Common information for all six scenarios is listed here. All the scenarios use the same spatial setup where the distance between the victim/disturber central and peripheral is around 10~cm, while the distance between the BLE victim and the disturber is around 20~cm, and no other Wi-Fi devices or BLE pairs are nearby that might interfere with the setup. The connection interval for both the victim and the disturber connections is set to 7.5~ms. The three scenarios and their difference are illustrated as follows:
\begin{enumerate}
	\item[(a1)] The CSAs are disabled in this scenario, and both the victim and the disturber connections are forced to communicate on the same BLE channel. There are two reasons behind this setting. First, it is used to simulate a rather harsh environment for the BLE communication, which implies that the whole 2.4~GHz frequency band is full of interference. Second, it provides easy-to-understand insights on the working principle of both models. The transmission power of the BLE victim devices in this scenario is 0~dBm while the BLE disturber is programmed to communicate using an output power of +8~dBm. The payload in each packet in the victim connection from both the central and the peripheral is designated to 50~bytes. The same payload size is also used for the disturber connection. The number of transactions in each connection interval of the disturber is 2, i.e. $x=2, n=4$. While the number of transactions in each victim connection interval increases from 1 to 5, i.e. $x=1\sim5, m=2\sim10$. It is worth mentioning that, although BLE specification does not limit the number of transactions in the connection interval, most BLE devices limit it to 5 due to reasons like limited hardware resources. However, it has been observed that most BLE devices do not exchange 5 pairs of packets during the communication. A possible explanation is the conflict between limited hardware resources, e.g., ram, and the scheduling of BLE stack. This scenario is considered as a preliminary validation of the proposed models.
	\item[(a2)] Similar to scenario~(a1), the CSAs are disabled in this scenario, and biased transmission powers are used, i.e. 0~dBm and +8~dBm. The number of transactions in each victim connection interval is fixed to 1. While for the disturber connection, it is 3. The payload of each packet inside the disturber connection is 50~bytes~($PT_D=512~\mu s$), and the payload of each victim packet raises from 100~bytes to 200~bytes, with a gap of 20~bytes~($PT_V=912~\mu s\sim1712~\mu s$). Due to the change of the disturber parameters, this scenario can be considered as a completely different electromagnetic environment from the first one. Hence, it is a further validation of the two models.
	\item[(a3)] All the parameters of scenario~(a3) are the same as the ones of scenario~(a2), except in this scenario CSAs are enabled. In this scenario, the CSA \#2 is enabled for both the victim and the disturber. Hence, scenario~(a3) can be considered a real-world use case. This scenario is considered as a realistic validation and evaluation of the accuracy of the models.
	\item[(b1)] Similar to scenario~(a1), the CSAs are disabled to simulate a rather harsh electromagnetic environment. The transmission power of the BLE disturber devices is decreased from +8~dBm to 0~dBm. So an unbiased transmission power is used for both the victim connection and the disturber one. This scenario is designed since it is common and realistic. There are many BLE devices using their default settings all the time. As a result, they communicate through a similar transmission power instead of biased ones. This scenario is considered a further validation of the models.
	\item[(b2)] The CSAs are still disabled in this scenario, and the unbiased transmission power (0~dBm) is used. All the other parameters follow the ones of scenario~(a2). Hence, this scenario again simulates an environment full of BLE connections and all of them use a same transmission power.
	\item[(b3)] All the parameters of scenario~(b3) are the same as the ones of scenario~(a3), except the transmission power. In this scenario, the transmission power of 0~dBm is implemented for both the victim connection and the disturber connection. Besides, the CSA \#2 is enabled, thus this scenario is another realistic use case. It is also the final validation and evaluation of the model.
\end{enumerate}

The experiments can be divided into experiment runs and sets. Each experiment run is a BLE connection with 1000 connection intervals. The throughput and the reliability are measured and calculated for the whole connection after 1000 connection intervals. In this way, these two performance metrics can reach a stable state under interference. Since it has been reported that a BLE connection requires a certain amount of time to reach a stable throughput and reliability when subjected to interference~\cite{pang_bluetooth_2022}. According to~\cite{pang_bluetooth_2022}, BLE communication performance metrics converge to their stable values after around 1000 connection intervals. As a result, 1000 connection intervals are planned for each experiment run. Each experiment run is repeated 500 times and forms an experiment set. Results are averaged over all experiment runs. This is to avoid a possible outlier from a single experiment run. As explained in~\cite{pang_novel_2023}, due to the varying connection event overlap probability and packet collision probability between BLE pairs, a single experiment run may lead to an extremely high or low result. Hence, multiple experiments runs are necessary to find the stable results. With the results shown in~\cite{pang_novel_2023}, 500 is chosen. As an example, scenario~(a1) increases its number of transactions from 1 to 5~($x=1\sim5$), hence, there are 5 experiment sets in total within the scenario. Extensive experiments are conducted just to ensure the correctness of the measured data, and to better validate the proposed models.

As mentioned in scenarios~(a1), (a2), (b1), and~(b2), the CSAs are disabled, and the BLE connections are designated to communicate on a single channel (channel 35). This channel is chosen since it is far away from popular Wi-Fi channels 1, 6, and 11, which are the major source of external interference in the office environment~\cite{wikipedia_list_2022}. To achieve this, Zephyr RTOS is deployed on the BLE development boards~\cite{zephyr_zephyr_2022}. Zephyr RTOS is an open-source real-time operating system for BLE that allows full control of the BLE stack. Link layer of BLE is modifiable so that multiple aspects of BLE communication, such as the CSAs and the number of transactions allowed within each connection event, can be manipulated~\cite{wikipedia_link_2022}.

\section{Results and Discussion}
\label{Results and Discussion}
In this section, the results from multiple experiments under the three designed scenarios are described first. They are used to validate the introduced throughput model and the reliability model. Two results are focused on, i.e. the reliability of the victim connection under interference, and the throughput of the victim connection under interference. As it will be shown, the results highlight the accuracy of both models. After the validation of both models, they are used to further discuss the trade-off in a BLE connection. Several Pareto curves between throughput and reliability are drawn, by varying different parameters in the models. They can be used as a design guideline for BLE developers/users. Meanwhile they indicate the advantage and convenience of the proposed models in both research and application domains of BLE.

\subsection{Validation}
As listed in Table~\ref{tab:table1}, most parameters can be set by the BLE connection itself, such as packet bit length~($L_V$) and connection interval~($CI_D$). However, the $BER$ is a parameter which needs careful attention, as it is not an input from the BLE connection, instead, an outcome of the electromagnetic environment. Hence, it is measured in this paper similar to the method mentioned in~\cite{pang_novel_2023, freschi_study_2019}. In general, packet corruption rate is first measured on the BLE victim connection. After that, the $BER$ is calculated by dividing the packet corruption rate by the bit length of the packet. This $BER$ is then used as an input for both models. It is worth noting that there are studies in estimating $BER$ based on the measured signal to noise ratio~\cite{wikipedia_signal--noise_2022}. However, to not introduce extra errors into the two developed models, the estimation of the $BER$ from signal to noise ratios is not considered.

\begin{figure}
	\centering
	\includegraphics[width=1\linewidth]{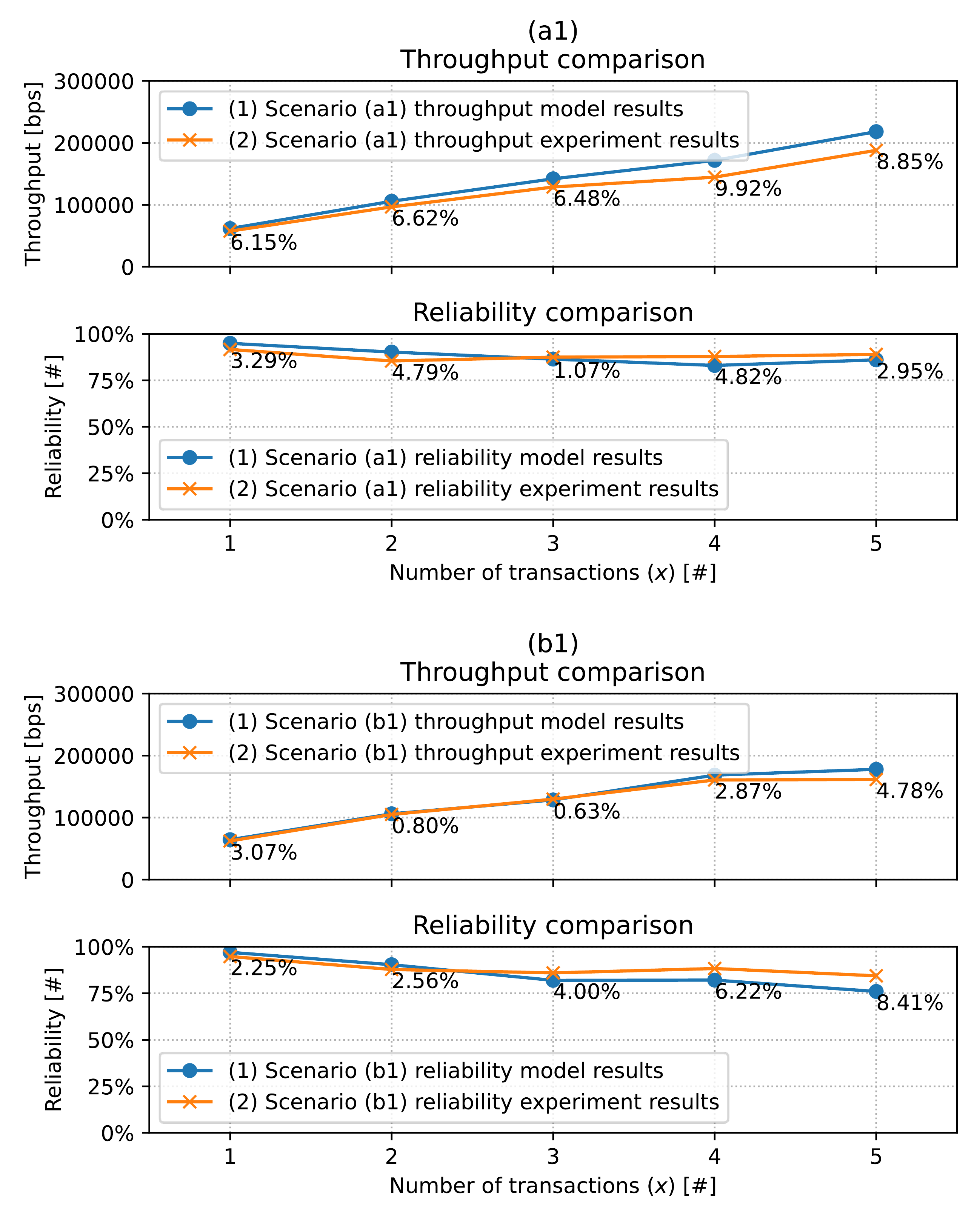}
	\caption{Comparison of experimental and model results under scenarios~(a1)~and~(b1). The deviations are displayed in percentage next to the curves. CSAs are disabled. Biased transmission powers (0~dBm and +8~dBm) are used in (a1), and an unbiased transmission power (0~dBm) is used in (b1). All the parameters are the same as described in experimental setup, hence, not repeated here. The predicted throughput and reliability are drawn as the blue lines with dots. The measured values of throughput and reliability are represented by the orange lines with crosses.}
	\label{fig:fig05}
\end{figure}
Fig.~\ref{fig:fig05} illustrates the validation results of scenarios~(a1)~and~(b1). For all the graphs in Fig.~\ref{fig:fig05}, the independent variable is the number of transactions within each victim connection event~($x$). As mentioned in the experimental setup, it increases from 1 to 5, i.e. $x=1\sim5,~m=2\sim10$. The deviations between the theory and the practice is shown as percentages next to the curves. Both the throughput and the reliability results in both scenarios show a clear correspondence between the models and their related experiments. In Fig.~\ref{fig:fig05}, the largest difference, 9.92\%, between the throughput model and its experiments appears when $x=4$ in scenario~(a1). The minimum difference between the model and the experiment is 0.63\% when $x=3$ in scenario~(b1). While the average difference from all the five data points is only 5.02\%. The largest error, 8.41\%, between the reliability model and its corresponding experiments shows when $x=5$ in scenario~(b1). The minimum difference is 1.07\%. The average difference is 4.04\%.

\begin{figure}
	\centering
	\includegraphics[width=1\linewidth]{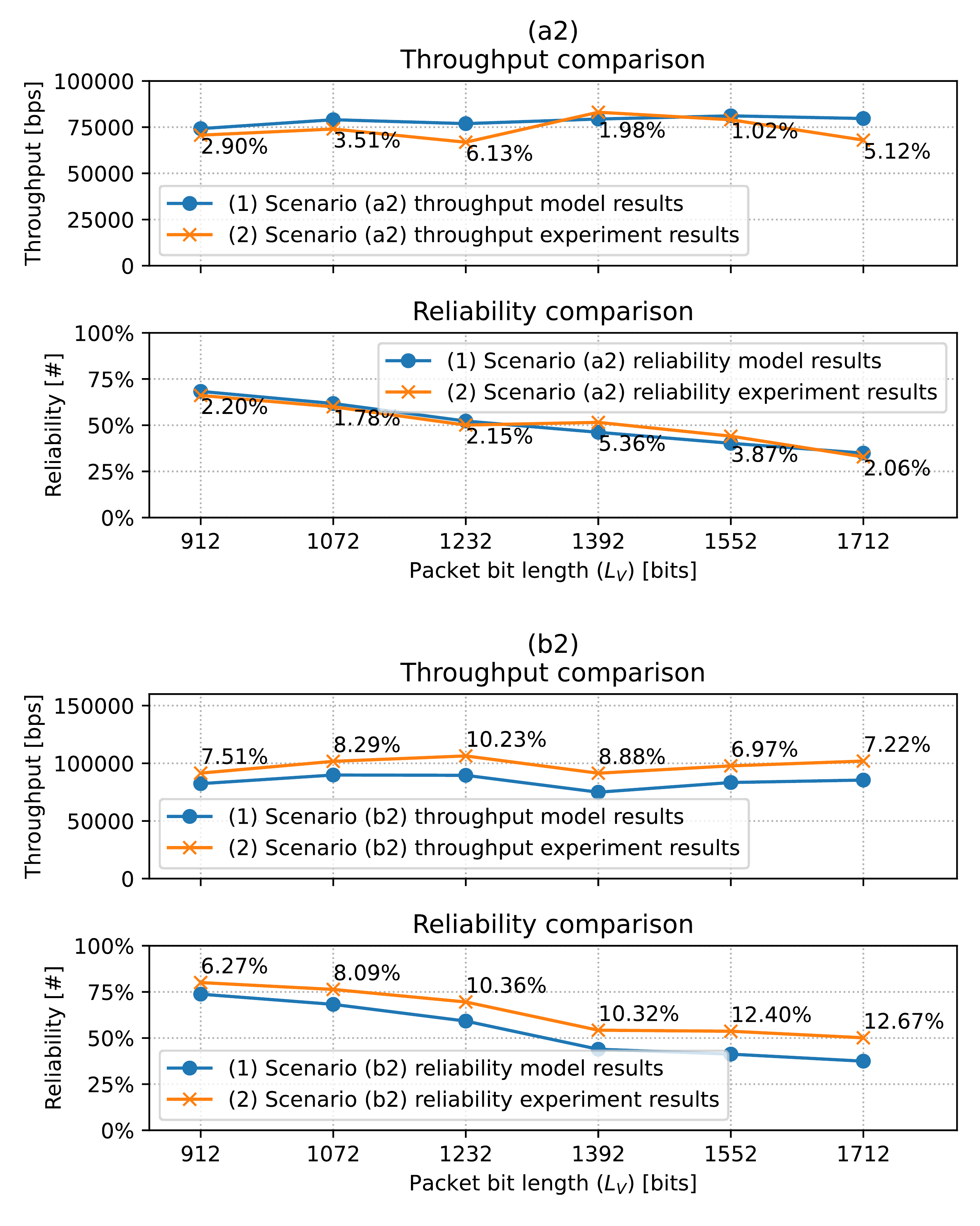}
	\caption{Comparison of experimental and model results under scenarios~(a2)~and~(b2). The deviations are displayed in percentage next to the curves. CSAs are disabled. Biased transmission powers (0~dBm and +8~dBm) are used in (a2), and an unbiased transmission power (0~dBm) is used in (b2). All the parameters are the same as described in experimental setup, hence, not repeated here. The predicted throughput and reliability are drawn as the blue lines with dots. The measured values of throughput and reliability are represented by the orange lines with crosses.}
	\label{fig:fig06}
\end{figure}
The validation results of scenarios~(a2)~and~(b2) are shown in Fig.~\ref{fig:fig06}. Different from Fig.~\ref{fig:fig05}, graphs in Fig.~\ref{fig:fig06} have the independent variable as the packet bit length in the victim connection~($L_V$). It increases from 912~bits to 1712~bits, corresponding to a payload size from 100~bytes to 200~bytes. The differences are displayed as a percentage value next to the curves. Similarly, the throughput and the reliability results show a consistency between the theory and the practice. In Fig.~\ref{fig:fig06}~(a2) throughput comparison, the largest error in the throughput validation is 6.13\%. This 6.13\% can be observed when $L_V=1232~bits$, i.e. the payload size of 140~bytes. The minimum and the average errors are observed as 1.02\% and 3.44\% respectively. In Fig.~\ref{fig:fig06}~(a2) reliability comparison, the largest reliability error between theory and practice is 5.36\%, when $L_V=1392~bits$, and the minimum value is 1.78\%. The average error in the reliability validation is only 2.91\%. In Fig.~\ref{fig:fig06}~(b2), the average deviation in the throughput validation is 8.18\%, while in the reliability comparison, the average difference is 10.02\%. Since they are the largest deviations across all experiments, they are regarded as potential deviation peaks for the two established models.

\begin{figure}
	\centering
	\includegraphics[width=1\linewidth]{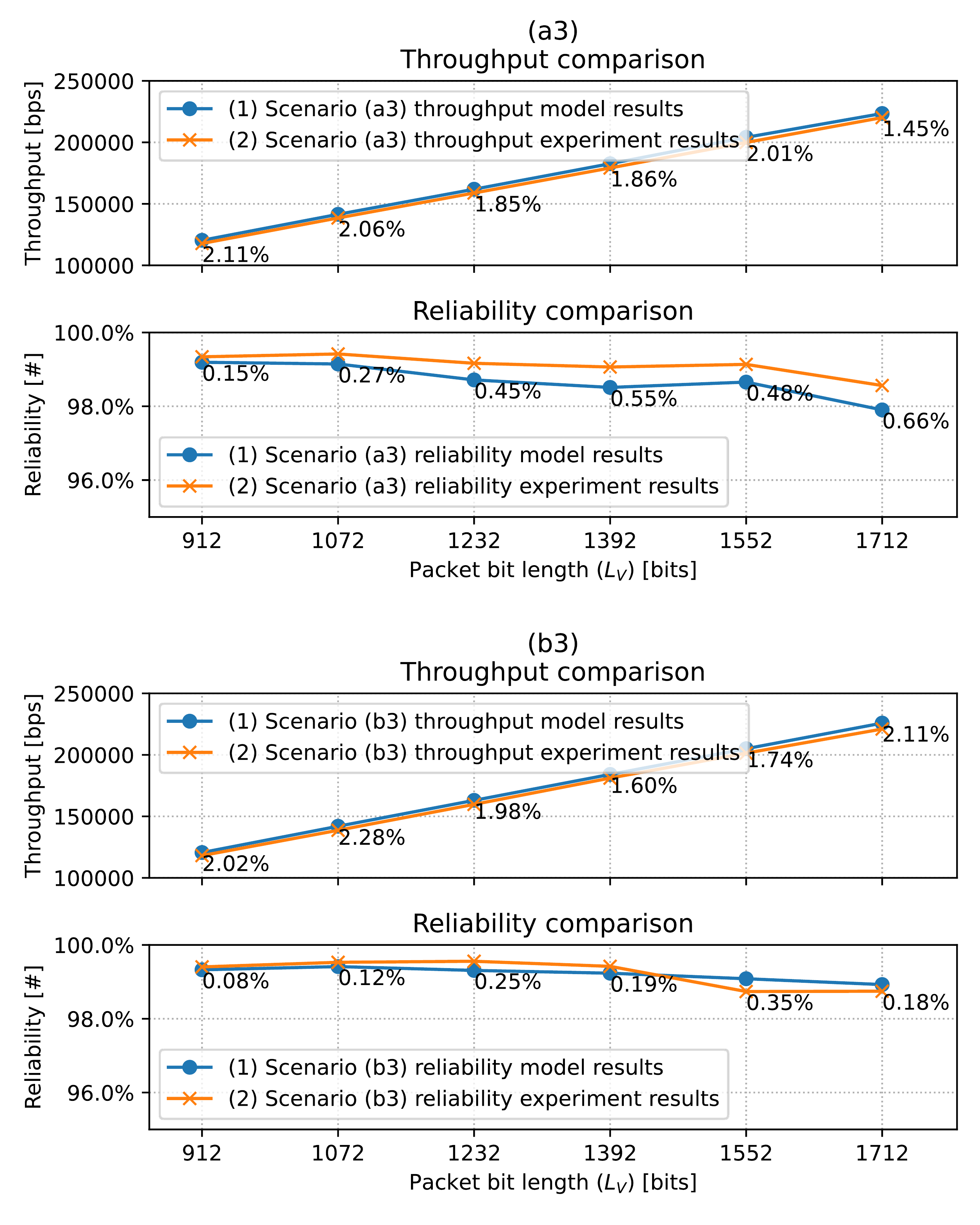}
	\caption{Comparison of experimental and model results under scenarios~(a3)~and~(b3). The deviations are displayed in percentage next to the curves. CSAs are enabled. Biased transmission powers (0~dBm and +8~dBm) are used in (a3), and an unbiased transmission power (0~dBm) is used in (b3). All the parameters are the same as described in experimental setup, hence, not repeated here. The predicted throughput and reliability are drawn as the blue lines with dots. The measured values of throughput and reliability are represented by the orange lines with crosses.}
	\label{fig:fig07}
\end{figure}
Fig.~\ref{fig:fig07} gives the results of the last validation experiment, scenarios~(a3)~and~(b3). Different from other scenarios, scenarios~(a3)~and~(b3) are considered real-world validations, since they strictly follows the BLE specification, including the use of CSAs. The independent variable is same as the one in scenarios~(a2)~and~(b2), which is the packet bit length in the victim connection~($L_V$). It ranges from 912~bits to 1712~bits, equaling a payload size of 100~bytes to 200~bytes. In Fig.~\ref{fig:fig07}~(a3), the maximum and minimum errors between the throughput model and the validation experiment are 2.11\% and 1.45\% respectively, and the average error is 1.89\%. As for the reliability validation in Fig.~\ref{fig:fig07}~(a3), the maximum and minimum differences are 0.66\% and 0.15\% only. The average difference is 0.43\%. In Fig.~\ref{fig:fig07}~(b3), the maximum and minimum deviations shown in the throughput comparison are 2.28\% and 1.60\%. The average value is 1.96\%. In the reliability comparison, the maximum, minimum, and average deviations are 0.35\%, 0.08\%, and 0.20\% respectively.

The results from all the six designed scenarios highlight the correspondence between the developed model and the practical experiments. However, it is worth mentioning that the results under various scenarios also illustrate some features of the developed models and BLE communications.

First, different accuracy is shown under different scenarios. It is typical that probability related experiments give varying results, e.g. the average differences from the throughput comparison of scenarios~(b1)~and~(b2) are 2.43\% and 8.18\% respectively. But these differences may also be from the BLE communication settings. For instance, a longer packet is more prone to be lost than a shorter packet under the same interference environment, according to Equation~\eqref{equation05}. Hence, the packets in scenario~(b2), 912~bits to 1712~bits, are more sensitive to interference than the packets in scenario~(b1), 512~bits. Furthermore, the interference in scenario~(b2) is stronger than the one in scenario~(b1) due to the increased number of transactions in the BLE disturber connection. Other possible factors include but are not limited to the unbiased transmission power and the possible coupling phenomenon between wireless signals and devices, which makes the communication more complicated and difficult to predict. These factors may lead to a higher chance for the BLE connection to be terminated. According to the BLE specification, the connection may be considered lost if it is experiencing continuous packet losses for six connection intervals~\cite{bluetooth_sig_bluetooth_2021}. So it is possible that a part of the deviations in scenario~(b2) is from the connection termination and reestablishment. Second, a lower error between the theoretical model and experimental results is shown in scenarios~(a3)~and~(b3), which are the realistic scenarios. In the other four scenarios (a1, b1, a2, and b2), the CSAs are disabled with the aim to simulate a harsh environment. The lower accuracy from those scenarios suggests that both developed models might provide lower prediction accuracy under harsh interference environments. It means that BLE communications are more difficult to be predicted under harsh environment. It is reasonable since BLE may have other actions under interference, such as connection termination and reestablishment. To further improve the accuracy of the model, an idea can be involving a mechanism for the BLE connection termination and reestablishment. However, extra research is necessary and thus it is considered as future work.

The differences between the theoretical models and the practical experiments can be briefly explained from two perspectives. First, due to the nature of probability, experimental outcomes are never exactly equal to but always converge toward theoretical values.~\cite{pages_numerical_2018}. Even after 500 experiment runs in each set, mostly the average of measured results can only fluctuate around the theoretical value. Second, it has been reported in literature that there are divergences between theory and measurement due to hardware differences and BLE stack implementation. Even under an environment without interference, the divergence can be up to 3\% to 6\%~\cite{ayoub_throughput_2020}.

\subsection{Trade-off Discussion}
The proposed models are used to analyze the trade-off between throughput and reliability of a BLE connection under interference after they have been validated under three different scenarios. The trade-off is illustrated in Pareto plots by varying some common parameters in the two models. Several instances are given below to better illustrate the use of the models and thus further discuss the trade-off within BLE communications.

\begin{figure}
	\centering
	\includegraphics[width=1\linewidth]{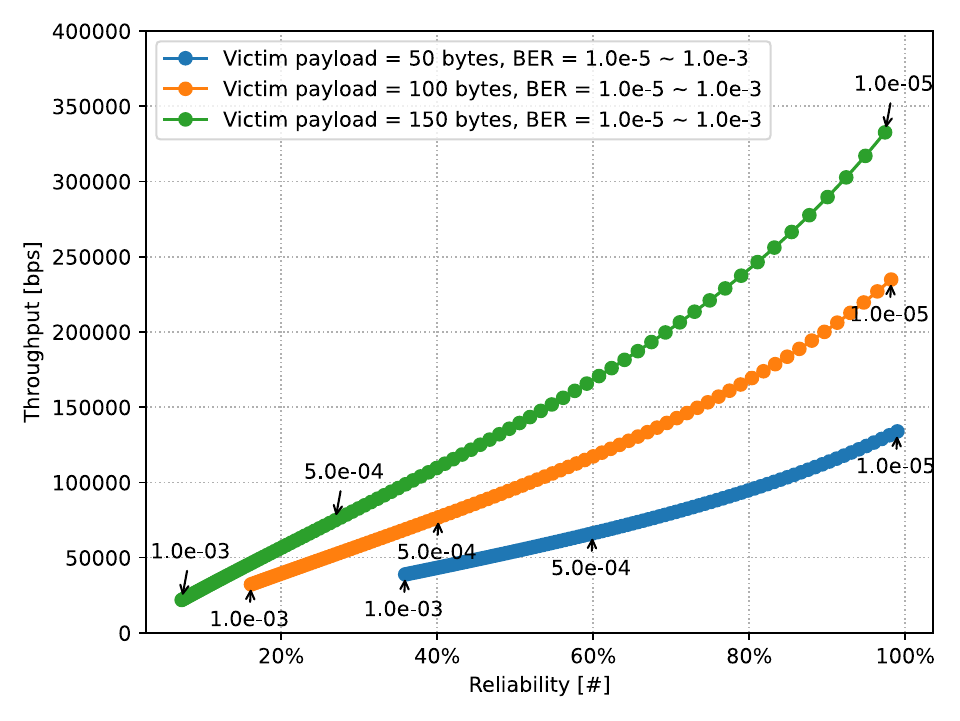}
	\caption{Pareto curves between BLE reliability and throughput when the BER varies. Parameters: $m$ = 4 ($x$ = 2), $n$ = 10, $PT_V$ = 512~$\mu$s, 912~$\mu$s, 1312~$\mu$s (payload = 50~bytes, 100~bytes, 150~bytes), $L_V$ = 512~bits, 912~bits, 1312~bits (payload = 50~bytes, 100~bytes, 150~bytes), $PT_D$ = 512~$\mu$s (payload = 50~bytes), $CI_V$ = 7.5~ms, $CI_D$ = 7.5~ms, $IFS$ = 150~$\mu$s, $BER$ = \mbox{1.0e-5} $\sim$ \mbox{1.0e-3}.}
	\label{fig:fig08}
\end{figure}
Fig.~\ref{fig:fig08} illustrates the compromise between BLE reliability and throughput when the environmental BER varies and all the other parameters stay the same. This can be linked to a use case where the distance between two BLE devices varies with time or activity. The BER variation ranges from a relatively low value of \mbox{1.0e-5} to the maximum sensitivity value of \mbox{1.0e-3} defined in BLE specification~\cite{gomez_overview_2012, bluetooth_sig_bluetooth_2021}. With the conditions described above, three Pareto curves between reliability and throughput are drawn based on different victim payload sizes, i.e. 50~bytes, 100~bytes, and 150~bytes. Any values can be used here, but these three values are chosen just to be examples. Taking the 50-byte payload size as an example, with an increment of the BER, the reliability of the BLE connection decreases from around 98\% to less than 40\%, while the throughput reduces from close to 150000~bps to less than 50000~bps. The decrement of both throughput and reliability is evidently nonlinear, which matches the results/conclusion reported in~\cite{pang_bluetooth_2022}.

This graph clarifies the impact of environment interference level on BLE communications when the connection settings of BLE are fixed. It also emphasizes the great influence of environments on BLE communications. For instance, when the BER equals to 1.0e-3, i.e., a harsh environment, the throughput from the three curves is on a similar level, but the reliability differs a lot from one another. A totally different case shows when the BER equals to 1.0e-5, i.e., a noiseless environment, the reliability of them is close to 100\%, but the throughput differs much. Besides, the graph also suggests that BLE communications should be carefully set in diverse environments. For example, under the harsh environment with a BER of 1.0e-3, it is better to send packets with a smaller payload size, e.g., 50~bytes, since it offers a higher reliability, and a similar throughput to the 150-byte payload size. When the interference level is in the middle, e.g., a BER of 5.0e-4, using different payload sizes does not lead to a large variation to the throughput, however, comparing with the case of 1.0e-3 BER, the throughput when the payload is 150~bytes is slightly higher than that when the payload is 50~bytes, although the reliability is much lower. With the interference level decreasing, the impact of environments also decreases, thus more potential of BLE communications is released. Hence, the throughput depends more on the BLE settings, e.g., payload size, instead of the environment.

\begin{figure}
	\centering
	\includegraphics[width=1\linewidth]{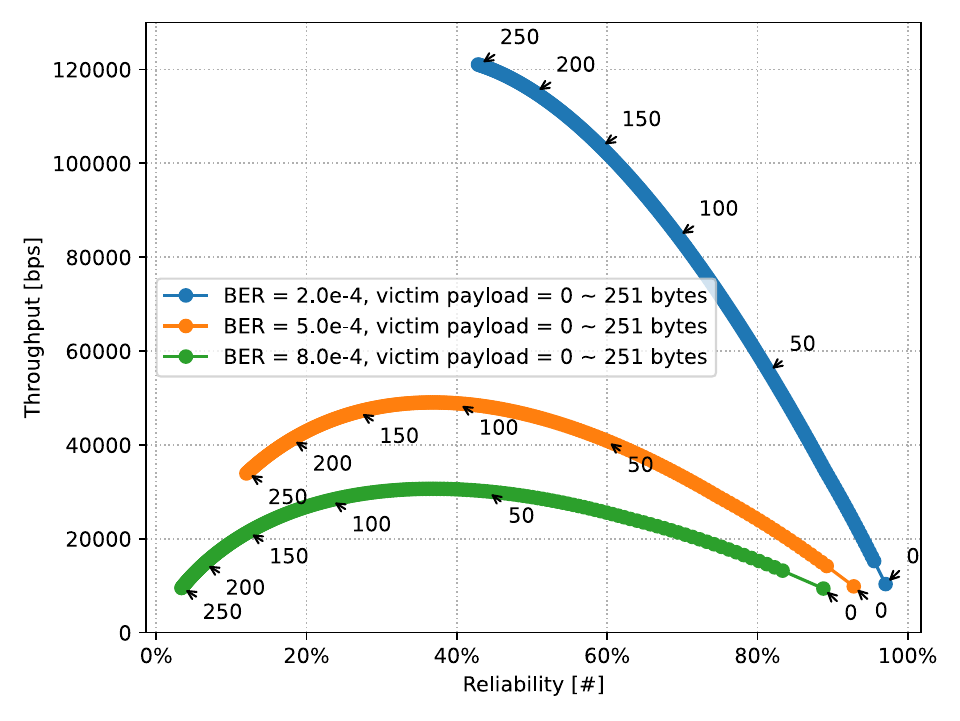}
	\caption{Pareto curve between BLE reliability and throughput when the payload in the victim connection varies. Parameters: $m$ = 2 ($x$ = 1), $n$ = 10, $PT_V$ = 80~$\mu$s $\sim$ 2120~$\mu$s (payload = 0~bytes $\sim$ 251~bytes), $L_V$ = 80~bits $\sim$ 2120~bits (payload = 0~bytes $\sim$ 251~bytes), $PT_D$ = 512~$\mu$s (payload = 50~bytes), $CI_V$ = 7.5~ms, $CI_D$ = 7.5~ms, $IFS$ = 150~$\mu$s, $BER$ = \mbox{2.0e-4}, \mbox{5.0e-4}, \mbox{8.0e-4}.}
	\label{fig:fig09}
\end{figure}
Fig.~\ref{fig:fig09} plots the trade-off while the payload size within the victim connection changes between 0 and 251~bytes, and all the other factors are fixed. This corresponds to a use case where a BLE connection adjusts its payload size frequently according to its application and need. The payload size of 0 to 251~bytes is the range defined by the BLE specification~\cite{bluetooth_sig_bluetooth_2021}. As a result, three Pareto curves are plotted by varying the BER (\mbox{2.0e-4}, \mbox{5.0e-4}, and \mbox{8.0e-4}). Similar to Fig.~\ref{fig:fig08}, we find no linear relationship between reliability and throughput. But, taking the BER of \mbox{5.0e-4} as an instance, there is a throughput peak displayed in Fig.~\ref{fig:fig09}. The peak point appears at the payload size of 120~bytes approximately, associated with a throughput of around 50000~bps and a reliability of 35\%. It suggests that the maximum throughput does not necessarily appear at the maximum reliability point. On the contrary, according to the curve of \mbox{5.0e-4} BER, the throughput reaches its lowest value when the BLE connection is most reliable. The reason behind it is that a smaller packet has a larger chance to be transferred successfully, but meanwhile with a shorter packet length, and hence, less throughput. However, when the payload size is over 120~bytes, both the reliability and the throughput begin to decrease. This is because the transmission success rate of a packet is too low. Despite the fact that each packet contains a large amount of data, few packets can be transmitted successfully. As a result, the throughput starts to decrease together with the reliability.

This graph nicely reveals the trade-off between the throughput and the reliability in BLE communications. Similar to Fig.~\ref{fig:fig08}, Fig.~\ref{fig:fig09} emphasizes the influence of environments. With BLE communications deployed in diverse environments, the interference levels from the environments differ the trade-off of BLE communications. In a less noisy environment, corresponding to the BER of 2.0e-4, the throughput increases and the reliability decreases, with the increment of the victim payload size. It is a similar phenomenon mentioned previously, the lower the interference level, the more potential of BLE communications can be released. With the environment becoming harsher and harsher, BLE settings should be adjusted so that the BLE communication reaches the requirements of the application. For instance, when the BLE communication is used for a localization application and the main requirement is to guarantee the continuity of the signal, a lower payload size should be used. With a lower payload size, the reliability can be maximized in any case, suggesting few packets are lost, and thus the continuity of the signal is guaranteed in any case. However, when the application expects the throughput to be the priority, the BLE communication must stop chasing the highest reliability, since that does not bring the highest throughput. Instead, increasing the payload size to a certain number according to the environment interference level can offer the throughput peak, with the sacrifice of some reliability. Another instance can be a BLE application with requirements on both latency and throughput under a harsh environment, e.g., with a BER of 8.0e-4. In this case, the BLE payload size should stay between 0 and around 50~bytes, instead of around 50 to 251~bytes. This is because the payload size between 50 and 251~bytes provides a similar throughput but a quite low reliability, thus the latency cannot be promised. Also due to the larger payload size, more energy might be wasted to transmit the packets, with the result of reaching the same throughput.

\begin{figure}
	\centering
	\includegraphics[width=1\linewidth]{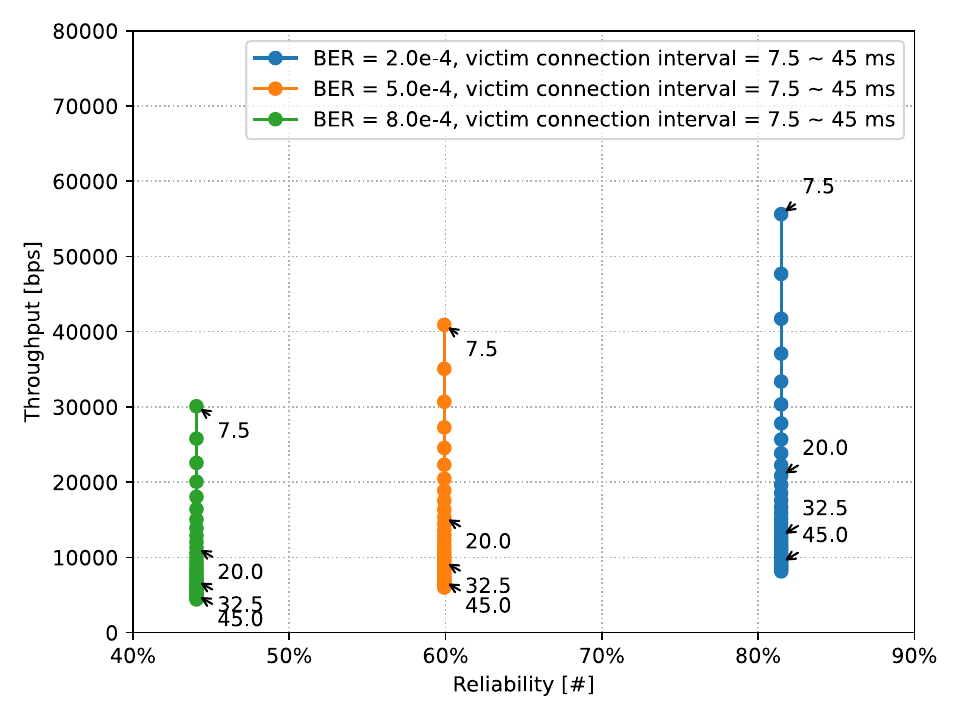}
	\caption{Pareto curve between BLE reliability and throughput when the connection interval in the victim connection varies. Parameters: $m$ = 2 ($x$ = 1), $n$ = 10, $PT_V$ = 512~$\mu$s (payload = 50~bytes), $L_V$ = 512~bits (payload = 50~bytes), $PT_D$ = 512~$\mu$s (payload = 50~bytes), $CI_V$ = 7.5~ms $\sim$ 45~ms, $CI_D$ = 7.5~ms, $IFS$ = 150~$\mu$s, $BER$ = \mbox{2.0e-4}, \mbox{5.0e-4}, \mbox{8.0e-4}.}
	\label{fig:fig10}
\end{figure}
As the last example, Fig.~\ref{fig:fig10} demonstrates the trade-off when the connection interval within the victim connection changes between 7.5 and 45~ms, and all the other factors are fixed. This can be related to a use case where a BLE connection adjusts its connection interval to limit the energy consumption. The connection interval starts with 7.5~ms since that is the lowest value allowed by the BLE specification~\cite{bluetooth_sig_bluetooth_2021}. The maximum value allowed by the specification is 4~s, however, the value of 45~ms is taken here just as illustration instance. Similar to previous examples, three Pareto curves are made by varying the BER (\mbox{2.0e-4}, \mbox{5.0e-4}, and \mbox{8.0e-4}). Three vertical lines are drawn as the relationship between throughput and reliability. They suggest that the reliability of the BLE victim connection is not impacted by its connection interval. This phenomenon can be explained by the reliability model, that is there is no victim connection interval involved in the model, hence, the victim connection interval does not influence the reliability of the BLE connection under the interference from other BLE connections. It is one of the conclusions in~\cite{pang_novel_2023}, and has been validated through experiments. With the reliability staying the same under the same BER, the BLE throughput decreases with the increment of the victim connection interval. It is reasonable, since the numerator of Equation~\eqref{equation03} stays the same, while its denominator increases.

The energy efficiency is not involved in this research, but according to existing literature, the larger the connection interval, the higher the energy efficiency~\cite{kindt_energy_2020}. Hence, due to no impact of the victim connection interval on the reliability, the connection interval can be adjusted to a low value as long as the throughput meets the requirement of the application. In this way, the energy efficiency of the BLE communication is improved. However, the adjustment of the value of the connection interval should also consider the latency requirement of the application. In one word, the connection interval can be decreased to improve the energy efficiency of the BLE communication while keeping the reliability the same, but other application requirements should also be considered, such as throughput and latency.

\section{Conclusion}
\label{Conclusion}
BLE is increasingly becoming a cornerstone of a variety of low-range IoT applications. The capability of organizing BLE communications in an efficient/effective manner under interference is crucial for successful deployments. Electromagnetic interference severely impacts the performance of a BLE connection, thus it is necessary to accurately model and experimentally evaluate BLE communications in realistic scenarios.

In this paper, we present three contributions: first, a mathematical model to estimate the throughput of a BLE connection subject to interference is derived and linked to the previously developed reliability model; second, extensive experiments on real-world BLE development boards are performed under different electromagnetic environments, and thus the proposed models and the combination of them are validated; third, the trade-off between BLE throughput and reliability is investigated through the validated models to illustrate some inside features of BLE communications.

The proposed throughput mathematical model is based on a Markov chain. Comparing with the state-of-the-art research in~\cite{dian_formulation_2020}, our model involves less parameters and calculations, and provides accurate results validated by practical experiments instead of just simulations. The combination of the BLE throughput and reliability gives a general idea on how to conduct the theoretical study when various BLE performance metrics are involved. The extensive experiments are meant to validate the correctness of the proposed models and assess their accuracy. Various scenarios are involved in the experiments, representing diverse electromagnetic environments, so the models are usable in different cases. With all the work above, we use the throughput and reliability models to investigate the trade-off within the BLE connection when subject to interference. To the best of our knowledge, this paper is the first one thoroughly quantifying the trade-off between BLE throughput and reliability. It can also be considered as the first step towards the smart management of large-scale BLE networks.

Regarding future work, three future research directions can be considered. First, other types of interference, such as Wi-Fi and ZigBee, might also be interesting to investigate. Normally, different interference types may cause some variations to the proposed mathematical models, hence, further research is necessary. Second, except throughput and reliability, there are many other performance aspects in BLE communications, such as energy efficiency. It would be interesting to see that other performance metrics can be developed as models as well, and combine them with one another to further study the insights of BLE communications. Third, a smart BLE network management system can be a promising research field. The current idea is to manage BLE networks through the developed models, and gradually update or optimize the models according to the use case, application, and environment. Another possible solution can be involving machine learning into the wireless communication management system, such as optimizing parameters in a BLE network management system.

% use section* for acknowledgment
%\section*{Acknowledgment}

%\begin{thebibliography}{1}
%\bibitem{IEEEhowto:kopka}
%H.~Kopka and P.~W. Daly, \emph{A Guide to \LaTeX}, 3rd~ed.\hskip 1em plus
%  0.5em minus 0.4em\relax Harlow, England: Addison-Wesley, 1999.
%\end{thebibliography}
\bibliographystyle{IEEEtranN}
\small
\bibliography{reference}

\end{document}